\documentclass[aps,pre,reprint,floatfix,amsfonts,amssymb,amsmath]{revtex4-1}
\usepackage[english]{babel}		
\usepackage[utf8]{inputenc}		
\usepackage{lmodern}			
\usepackage[T1]{fontenc}		
\usepackage{xcolor}			
\usepackage{mathtools}			
\usepackage{bm}				
\usepackage{graphicx}			
\usepackage{enumerate}			
\usepackage{booktabs}			
\usepackage{multirow}			
\usepackage[]{units} 			
\usepackage{braket} 			
\usepackage{url}			
\usepackage[colorlinks=true,linkcolor=uniDarkBlue,citecolor=uniDarkBlue,urlcolor=uniDarkBlue,breaklinks=true]{hyperref}
\usepackage[capitalize]{cleveref}	

\definecolor{uniAntra}{HTML}{323232}
\definecolor{uniWhite}{HTML}{FFFFFF}
\definecolor{uniLightBlue}{HTML}{00BEFF}
\definecolor{uniDarkBlue}{HTML}{004191}
\definecolor{moreland1}{HTML}{3b4cc0}
\definecolor{moreland8}{HTML}{b40426}

\newcommand{\A}{\mathrm{a}}

\newcommand{\cc}{\mathrm{c}}
\newcommand{\dd}{\mathrm{d}}
\newcommand{\dt}{\mathrm{d}t}
\newcommand{\dm}{\varDelta\mu}
\newcommand{\ee}{\mathrm{e}}
\newcommand{\ff}{\mathrm{f}}
\newcommand{\ii}{\mathrm{i}}

\newcommand{\PP}{\mathrm{P}}
\newcommand{\st}{\mathrm{ss}}
\newcommand{\GDP}{\mathrm{GDP}}
\newcommand{\GTP}{\mathrm{GTP}}
\newcommand{\kb}{k_\mathrm{B}}
\newcommand{\re}{\operatorname{Re}}

\begin{document}



\title{Interlinked GTPase cascades provide a motif for both robust switches and oscillators} 
\author{Andreas Ehrmann}
\author{Basile Nguyen}
\author{Udo Seifert}
\email[For correspondence: ]{useifert@theo2.physik.uni-stuttgart.de}
\affiliation{II. Institut für Theoretische Physik, Universität Stuttgart, 70550 Stuttgart, Germany}
\date{\today}
\begin{abstract}
 GTPases regulate a wide range of cellular processes, such as intracellular vesicular transport, signal transduction, and protein translation. These hydrolase enzymes operate as biochemical switches by toggling between an active guanosine triphosphate (GTP)-bound state and an inactive guanosine diphosphate (GDP)-bound state. We compare two network motifs, a single-species switch and an interlinked cascade that consists of two species coupled through positive and negative feedback loops. We find that interlinked cascades are closer to the ideal all-or-none switch and are more robust against fluctuating signals. While the single-species switch can only achieve bistability, interlinked cascades can be converted into oscillators by tuning the cofactor concentrations, which catalyse the activity of the cascade. These regimes can only be achieved with sufficient chemical driving provided by GTP hydrolysis.  In this study, we present a thermodynamically consistent model that can achieve bistability and oscillations with the same feedback motif.
\end{abstract}


\maketitle


\section*{Introduction}\label{sec:intro}
Interlinked cascades are a special class of network motifs that are fundamental components of cell signalling networks. They regulate the cellular response to an input signal using complex feedback structures and involve a multitude of proteins. Ideally, these molecular switches should have an all-or-none response, thus making them ultrasensitive \cite{tuyu08,ferr14}. Although this feature is essential for signal transducers, their steady state output is by definition sensitive to input fluctuations. Molecular switches can benefit from bistability which makes them robust against input noise \cite{ferr14}. Interlinked cascades often consist of two species coupled through positive and negative feedback loops. Positive feedback loops can enhance the input signal as well as the bistability region, and make the switch more robust against fluctuations. Negative feedback loops are necessary for oscillations and enable a fast input response. The combination of positive and negative feedback is a highly versatile system as it facilitates rich dynamics such as bistability or oscillations, and is robust against input noise \cite{pfeu09,kim06,kim08,koch17}.

Bistability can already be achieved by a minimal model containing only one species and an autocatalytic reaction as in Schlögl's model \cite{schl72,vell09}, complex feedback mechanisms are not required. A natural question that arises is why interlinked cascades should be advantageous over single-species switches. In this study, we show that the interlinked cascade has an improved switching quality, which is closer to the ideal all-or-none switch compared to its single-species counterpart. Moreover, interlinked cascades can decrease the effect of input fluctuations \cite{hoos05,zhan07}.

In this paper, we will focus on small GTPase proteins, which are key players in many cellular processes such as gene expression, membrane trafficking or vesicular transport \cite{witt14a,witt14b,mizu12}. Their structure comprises three domains: a nucleotide-binding domain, where a guanosine triphosphate (GTP) or guanosine diphosphate (GDP) is lodged, a catalytic-binding domain, and an effector-binding domain \cite{witt14a,witt14b}. The effector-binding domain is responsible for different functions in membrane transport, such as vesicle budding, movement, tethering, and fusion \cite{behn05,huta11}. Small GTPases have a high affinity to nucleotides and are mostly found in complexes with GTP or GDP under physiological conditions. In the GTP-bound state, small GTPases are in an open activated state and can selectively bind to a specific set of effectors with high affinity. In the GDP-bound state, small GTPases are preferably closed and the system is considered inactivated as they have a low affinity to effectors.

Small GTPases rely on cofactors to regulate their activity. Cofactors are essential catalysers since the spontaneous GTPase rate is very low without them ($\unit[10^{-4}]{s^{-1}}$). Small GTPases can on the one hand form a complex with GTPase-activating proteins (GAPs), which greatly accelerate the GTPase rate (deactivation) up to $\unit[10]{s^{-1}}$ \cite{gide92}. On the other hand, they can bind to guanine nucleotide exchange factors (GEFs) that catalyse the dissociation of GDP in the inactive state (up to $10^5$-fold \cite{kleb95}), followed by an activation with GTP. Hence, cofactors are essential to realise fast and reliable switching behaviour. In this paper, we show that these cofactor concentrations must be tuned appropriately to achieve bistability or oscillations. Recently, it was proposed that small GTPases are optimised to rapidly bind to their effectors and quickly release them \cite{nguy17}, which is ideal for tasks like defining identities within the membrane \cite{barr13}. Defects or mutations in the cofactors' functions have been experimentally linked to reduced targeting \cite{bl13,cher13}. Small GTPases share many similarities with kinetic proofreading \cite{hopf74,nini75,benn79}, Min proteins \cite{huan03,loos08,hala12,xion15,wu16,denk18} and the $\unit[70]{kDa}$ heat shock protein (Hsp70), which has an analogous ATP cycle and set of cofactors \cite{yuji13,fink16,rios14,nguy17}. Recent experiments have shown that the Hsp70-Hsp90 cascade can improve client folding yields \cite{mora18}.

\begin{figure}\centering
  \includegraphics[width=0.48\textwidth]{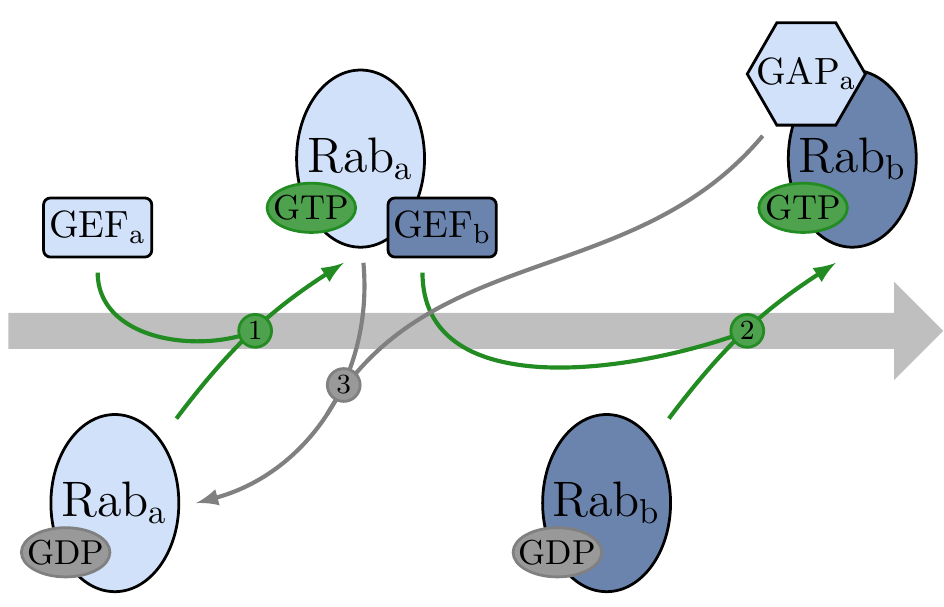}
 \caption{Rab GEF and GAP cascade. In step 1, the activation of the first Rab protein is catalysed by GEF$_\A$. In their active form, Rab proteins can bind to the GEF of the upstream Rab, which catalyses its activation as shown in step 2. If a GAP binds to an active Rab, the inactivation of the downstream Rab is catalysed after it has fulfilled its function as marked by step 3. The overall behaviour is a transition from left to right as indicated by the grey arrow. Figure adapted from \cite{huta11}.}
 \label{fig:casade}
\end{figure}

Small GTPases from different families are linked through feedback schemes to accomplish either bistability or oscillations. Interlinked cascades are designed to achieve bistability, they are found in many parts of the cell, such as the Rab5-Rab7 in the endosome or Rab/Ypt in the secretory pathway \cite{orti02,pusa12,rive09,nott11,suda13,rana15}. While each species has its specific set of cofactors and effectors, all of them share the same structure. The effector of the first species is the GEF of the second species, which catalyses the activation of the second system (GEF cascade). Once activated, the second species can bind to its effector and the GAP of the first species. When bound to the latter, it inactivates the first system (GAP cascade) as illustrated in \cref{fig:casade}. Errors when deactivating the previous species have been linked to trafficking defects \cite{rive09,nott11,suda13,rana15}. Interestingly, small GTPases are also tuned to achieve oscillations. For example, Rho GTPases are drivers of cellular morphogenesis and serve as pattern to guide the formation of various cellular structures \cite{gory08,beme15,gory16,grae17}. In this paper, we show that interlinked small GTPases can be tuned to act as molecular switches or oscillators with the same feedback structure, thus providing a versatile motif. Fischer-Friedrich {\sl et al.} have shown experimentally that the MinDE system, which shares a similar feedback structure, displays a transition from a switch to an oscillator as the cell length increases \cite{fisc10}.

Small GTPases are GTP-driven machines that rely on the GTP hydrolysis cycle, which transforms a GTP into a GDP and an inorganic phosphate $\PP_\ii$. Under physiological conditions, an excess of GTP is maintained, driving the system far from equilibrium \cite{trau94}. Thermodynamics states that complex behaviour such as bistability or oscillations can only be achieved far from equilibrium \cite{qian05,ge09,xion15,bara17,nguy18,herp18}. Although extensive research has been carried out on small GTPase cascades, few studies have investigated the finite energy budget of the hydrolysis cycle. Apart from thermodynamically consistent models for a phosphorylation-dephosphorylation switch \cite{qian05,ge09,bish10,zhuh10,ge11}, very little is known about the thermodynamic properties of interlinked cascades. Thermodynamic consistency requires mass action kinetics, thus forbidding more complex activation functions such as sigmoidal, Michaelis-Menten, and Hill functions, which are primarily used in the literature to analyse interlinked cascades \cite{stad77,choc77,gard00,mark04, delc08,liud11,tiwa12,huan16}. Moreover, the underlying biological reactions of these activations remain unclear in contrast to the fundamental reactions considered in the present study.


\section*{Model description}\label{sec:mat+meth}


\subsection*{Thermodynamically consistent model for single-species switches}

\begin{figure}\centering
  \includegraphics[width=0.25\textwidth]{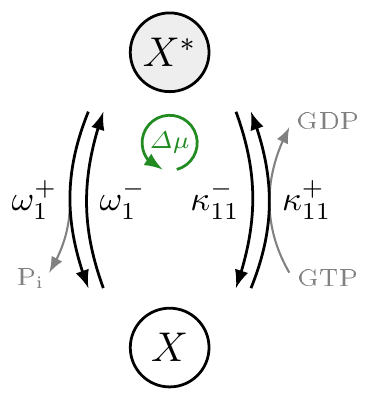}
 \caption{Model for the single-species switch. Transitions from the inactive state $X$ to the active state $X^*$ mediated by the hydrolysis and nucleotide exchange reactions of \cref{eq:reacts1w,eq:reacts1k} with the associated transition rates $\omega_1^\pm$ and $\kappa_{11}^\pm$. Completing the anticlockwise cycle corresponds to the hydrolysis of one GTP, which liberates a free energy $\dm$.}
 \label{fig:feedback}
\end{figure}

We consider a biological switch that consists of species $X$ in a volume $\Omega$. It can toggle between an inactive GDP-bound state $X$ and an active GTP-bound state $X^*$. Transitions between these two states occur through a hydrolysis reaction
\begin{equation}
 X^* \xrightleftharpoons[\omega_1^-]{\omega_1^+} X + \PP_\ii,
 \label{eq:reacts1w}
\end{equation}
and a nucleotide exchange reaction
\begin{equation}
 I + 2X^* + X + \GTP \xrightleftharpoons[\kappa_{11}^-]{\kappa_{11}^+} I + 3X^* + \GDP,
 \label{eq:reacts1k}
\end{equation}
where $\omega_1^\pm, \kappa_{11}^\pm$ are transition rates shown in \cref{fig:feedback}(a). We assume that the input $I$ acts as a catalyst by boosting the nucleotide exchange reactions like the GEF molecules for small GTPases. Second, we suppose that two active proteins are required to activate a third one and write $\kappa_{11}^\pm$ with double indices to indicate this self-feedback. We choose this effective reaction in order to achieve non-linear feedback, which is a necessary condition for bistability. In reality, a GEF only binds to one active protein to activate a second one \cite{kleb95b,zhuh10,stan16}. This would normally result in one $X^*$ term on the left side of \cref{eq:reacts1k}. However, Goryachev {\sl et al.} \cite{gory08} have shown that a model for Rho GTPases with membrane interaction containing only bimolecular reactions can be simplified to an effective trimolecular reaction similar to \cref{eq:reacts1k}. Note that we did not consider the equilibrium binding reactions in our model, they can be coarse-grained in a thermodynamically consistent way assuming timescale separation \cite{espo12,wach18}. 

A transition from $X$ to $X^*$ along the ``+'' direction of reaction \eqref{eq:reacts1k} and back to $X$ along the ``+'' direction of reaction \eqref{eq:reacts1w} does not change the state of the system, however, it turns one GTP into one GDP and a P$_\ii$. Such a complete cycle consumes a chemical energy
\begin{equation}
 \dm \equiv \mu_\GTP^{} - \mu_\GDP^{} - \mu_{\PP_\ii}^{} = \dm_0 + \kb T\ln\frac{[\GTP]}{[\GDP][\PP_\ii]},
 \label{eq:dmu_def}
\end{equation}
where $\mu_S^{}$ is the chemical potential of species $S=\GTP,\GDP,\PP_\ii$ and $[S]$ its concentration. The temperature is denoted by $T$ and $\kb$ represents Boltzmann's constant.
The last equality is the approximation for an ideal solution, where $\dm_0$ is a reference value. Physiological conditions are maintained by an excess of GTP, characterised by $\dm>0$, and equilibrium corresponds to $\dm=0$. The transition rates must satisfy the local-detailed balance condition \cite{seif11}, which reads
\begin{equation}
 \frac{\kappa_{11}^+\omega_1^+}{\kappa_{11}^-\omega_1^-} = \ee^{\dm/\kb T}.
 \label{eq:ldb_one}
\end{equation}


The state of the chemical reaction system \eqref{eq:reacts1w} and \eqref{eq:reacts1k} is determined by only one variable, the number of active proteins $n$. The time evolution of the probability distribution $P(n,t)$ to find the system in state $n$ at time $t$ is governed by the chemical master equation (CME), which reads
\begin{equation}
\begin{aligned}
 \partial_t P(n,t) &= \alpha_{n-1}P(n-1,t) + \beta_{n+1}P(n+1,t) \\
 &- (\alpha_n+\beta_n)P(n,t).
 \label{eq:cme_one}
 \end{aligned}
\end{equation}
Here, we define rate parameters as
\begin{equation}
\begin{aligned}
 \alpha_n = \alpha_n^\kappa + \alpha_n^\omega &\equiv \frac{\kappa_{11}^+}{\Omega^2}(N-n)n(n-1)[I] + \omega_1^-(N-n), \\
 \beta_n = \beta_n^\kappa + \beta_n^\omega &\equiv \frac{\kappa_{11}^-}{\Omega^2}n(n-1)(n-2)[I] + \omega_1^+n,
 \label{eq:ratepara_one}
\end{aligned}
\end{equation}
and introduce the total number of proteins $N$ as well as the input concentration $[I]$, which is given by the number of $I$ particles divided by the volume $\Omega$.
For long times $t$ and constant $[I]$, the system reaches a non-equilibrium steady state (NESS), at which the left-hand side of \cref{eq:cme_one} vanishes, leading to the condition $\alpha_{n-1}P_\st(n-1)=\beta_nP_\st(n)$. The analytic solution for the steady state probability distribution reads
\begin{equation}
\begin{aligned}
 \frac{P_\st(n)}{P_\st(0)} = \prod_{i=0}^{n-1} \frac{\alpha_i}{\beta_{i+1}}, && P_\st(0)=1-\sum_{j=1}^\infty P_\st(j).
 \label{eq:prob_one}
 \end{aligned}
\end{equation}

From the CME \eqref{eq:cme_one}, we obtain the equation for the time evolution of the concentration of active proteins
\begin{equation} 
 x^* \equiv \frac{1}{\Omega}\sum_{n=0}^N nP(n),
 \label{eq:concen_def}
\end{equation}
which reads in the deterministic limit ($N \to \infty,\Omega \to \infty$)
\begin{equation}
\begin{aligned}
 \frac{\dd x^*}{\dt} &= \left(\omega_1^- + [I]\kappa_{11}^+{x^*}^2 \right)(1-x^*) -\left(\omega_1^+ + [I]\kappa_{11}^-{x^*}^2\right)x^*.
 \label{eq:det-one}
 \end{aligned}
\end{equation}
As the total number of proteins $N$ is conserved, we set the sum of active and inactive concentrations to one (i.e. $x+x^*=1$), thus, setting $N/\Omega=1$.
Note that the steady state concentration for zero input is simply given by $x^*([I]{=}0)=\omega_1^-/(\omega_1^++\omega_1^-)$.
Having introduced the model for single-species switches, we consider interlinked cascades of different species $X,Y$ with positive self-feedback in the next section.



\subsection*{Thermodynamically consistent model for interlinked cascades}

\begin{figure*}\centering
\includegraphics[width=\textwidth]{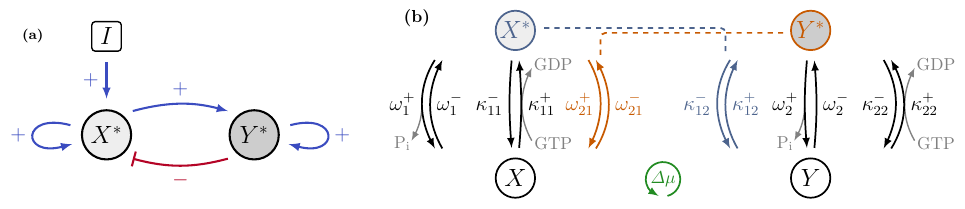}
 \caption{Model for the interlinked cascade. (a) Interlinked positive and negative feedback loops between the active forms $X^*$ and $Y^*$ of the two species. Blue arrows represent activation and the red line represents inhibition. The input $I$ has a direct effect on species $X$. (b) Transitions from the inactive states $X,Y$ to the active states $X^*,Y^*$ mediated by the hydrolysis and nucleotide exchange reactions of \cref{eq:two_reacts}. The feedback between $X$ and $Y$ is illustrated in blue and orange. Completing the anticlockwise cycles corresponds to the hydrolysis of one GTP, which liberates a free energy $\dm$.}
 \label{fig:two_fb+reacts}
\end{figure*}

As in the model for the single-species switch, we assume that the input $I$ only affects species $X$ and catalyses its activation. We now introduce a positive feedback from the first $X^*$ to the second $Y^*$ component of the cascade and negative feedback from $Y^*$ to $X^*$, as shown in \cref{fig:two_fb+reacts}(a). The underlying biochemical reactions of the interlinked cascade are the following set of hydrolysis and nucleotide exchange reactions
\begin{equation}
\begin{aligned}
  X^* &\xrightleftharpoons[\omega_1^-]{\omega_1^+} X + \PP_\ii, \\
  I + 2X^* + X + \GTP &\xrightleftharpoons[\kappa_{11}^-]{\kappa_{11}^+} I + 3X^* + \GDP, \\
  G_2^+ + 2X^* + Y + \GTP &\xrightleftharpoons[\kappa_{12}^-]{\kappa_{12}^+} G_2^+ + 2X^* + Y^* + \GDP, \\
  Y^* &\xrightleftharpoons[\omega_2^-]{\omega_2^+} Y + \PP_\ii, \\
  G_2^+ + 2Y^* + Y + \GTP &\xrightleftharpoons[\kappa_{22}^-]{\kappa_{22}^+} G_2^+ + 3Y^* + \GDP, \\
  G_1^- + 2Y^* + X^* &\xrightleftharpoons[\omega_{21}^-]{\omega_{21}^+} G_1^- + 2Y^* + X + \PP_\ii,
  \label{eq:two_reacts}
  \end{aligned}
\end{equation}
which are illustrated in \cref{fig:two_fb+reacts}(b).
Note that the first two reactions of \cref{eq:two_reacts} are the same as \cref{eq:reacts1w,eq:reacts1k} of the single-species switch and that the additional feedback pathways are realised by the remaining four reactions. The molecules $G_1^-$ and $G_2^+$ can boost the hydrolysis and nucleotide exchange reactions, respectively, similar to the input $I$.
In the case of small GTPases, $G_1^-$ would correspond to the GAP of species $X$ and $G_2^+$ to the GEF of species $Y$.
Local-detailed balance requires
\begin{equation}
\begin{aligned}
 \frac{\kappa_{11}^+\omega_1^+}{\kappa_{11}^-\omega_1^-} &= \ee^{\dm/\kb T}, & \frac{\kappa_{22}^+\omega_2^+}{\kappa_{22}^-\omega_2^-} &= \ee^{\dm/\kb T}, \\
 \frac{\kappa_{12}^+\omega_2^+}{\kappa_{12}^-\omega_2^-} &= \ee^{\dm/\kb T}, & \frac{\kappa_{11}^+\omega_{21}^+}{\kappa_{11}^-\omega_{21}^-} &= \ee^{\dm/\kb T}.
 \label{eq:ldb_two}
\end{aligned}
\end{equation}
The deterministic equations for the normalised active protein concentrations $x^*$ and $y^*$ are given by the following set of non-linear ordinary differential equations (ODEs)
\begin{equation}
\begin{aligned}
 \frac{\dd x^*}{\dt} &= \left(\omega_1^- + [I]\kappa_{11}^+{x^*}^2 + [G_1^-]\omega_{21}^- {y^*}^2\right)(1-x^*) \\&\quad- \left(\omega_1^+ + [I]\kappa_{11}^-{x^*}^2 + [G_1^-]\omega_{21}^+ {y^*}^2\right)x^*, \\
 \frac{\dd y^*}{\dt} &= \left(\omega_2^- + [G_2^+]\kappa_{22}^+{y^*}^2 + [G_2^+]\kappa_{12}^+{x^*}^2\right)(1-y^*) \\&\quad- \left(\omega_2^+ + [G_2^+]\kappa_{22}^-{y^*}^2 + [G_2^+]\kappa_{12}^-{x^*}^2\right)y^*.
 \label{eq:det-two}
\end{aligned}
\end{equation}
The results presented in the next section do not depend strongly on the choice of parameters. The explicit parameter values used in the simulations can be found in the Methods section.


\section*{Results}\label{sec:res}


\subsection*{Bistability analysis of single-species switches}

\subsubsection*{Deterministic analysis}

\begin{figure*}\centering
\includegraphics[width=\textwidth]{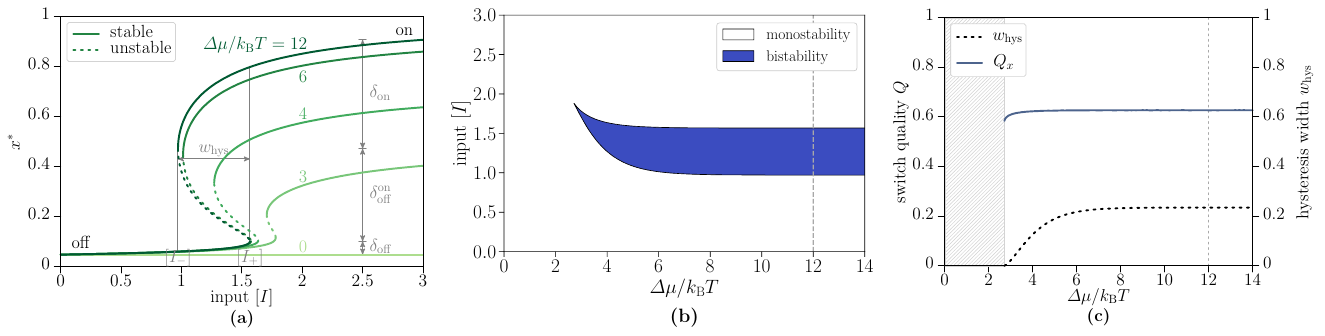}
 \caption{Characteristics of the single-species switch. (a) Active concentration $x^*$ as function of the input $[I]$ for different chemical affinities $\dm/\kb T=0,3,4,6,12$. A higher chemical affinity enables bistability and an activation for which smaller input $[I]$ is sufficient. (b) Phase diagram of the single-species switch. The bistability region arises for a critical value of the chemical affinity $\dm_\cc/\kb T\simeq2.72$. (c) Relative hysteresis width $w_\text{hys}$ from \cref{eq:w_hys} and switch quality $Q_x$ from \cref{eq:def_quality} as functions of the chemical affinity $\dm$. The switch quality of the single-species switch is $Q_x\simeq0.63$ at $\dm/\kb T=12$ and the associated hysteresis width is $w_\text{hys}\simeq0.235$. The hysteresis width and the activation degrees are marked in (a) for $\dm/\kb T=12$.}
 \label{fig:act+bi_one}
\end{figure*}

We first consider the deterministic \cref{eq:det-one} and study its non-equilibrium steady state solutions for which the left-hand side vanishes. Depending on the input concentration $[I]$ and the chemical affinity $\dm$, either one or three real solutions $x_i^*$, with $i=-,0,+$, are possible. Hence, one or three different steady states can occur as shown in \cref{fig:act+bi_one}(a). The solid lines correspond to stable steady states, the dashed lines to unstable steady states. If the chemical affinity is large enough, two stable steady states coexist and the system can exhibit hysteresis, which is a characteristic of bistability. It is interesting to note that an increase of the chemical affinity $\dm$ enables an activation for which smaller input $[I]$ is sufficient. Bistability can only occur for non-vanishing chemical affinity, i.e., under non-equilibrium conditions. In \cref{fig:act+bi_one}(b), we show the phase diagram, where the chemical affinity and the input concentration are varied. The bistability region arises for a critical value of the chemical affinity $\dm_\cc/\kb T\simeq2.72$ and increases for larger affinities.

We analyse the bistability region in more detail by introducing the relative width of the hysteresis zone \cite{qian05}
\begin{equation}
 w_\text{hys} = \frac{[I_+]-[I_-]}{[I_+]+[I_-]},
 \label{eq:w_hys}
\end{equation}
with upper $[I_+]$ and lower $[I_-]$ bounds of the hysteresis region for a given chemical affinity $\dm$. As shown in \cref{fig:act+bi_one}(c), the minimal chemical affinity for hysteresis is $\dm_\cc/\kb T\simeq2.72$. For values $\dm>\dm_\cc$, the relative hysteresis width growths until it saturates at its plateau value $w_\text{hys}\simeq0.235$ for $\dm/\kb T=12$.
From a theoretical perspective, a large hysteresis width has its assets and drawbacks. The major advantage is its increased robustness against input fluctuations, which is very beneficial in biological systems. However, it slows down the activation process because more input is needed for the entire activation in comparison with monostable activation functions like sigmoids. In this sense, an all-or-none activation with maximum distinction between the inactive and active state together with an appropriate hysteresis width is desirable.

Hence, as a measure of the performance of biochemical switches, we define the switch quality
\begin{equation}
 0 \leq Q \equiv \big((1-\delta_\text{off}) + (1-\delta_\text{on}) + \delta_\text{off}^\text{on}\big)/3 \leq1,
 \label{eq:def_quality}
\end{equation}
with the activation degrees in the ``off'' and ``on'' state as well as in the transition from ``off'' to ``on'' as
\begin{equation}
\begin{aligned}
 \delta_\text{off}(x^*) \equiv \max_\text{off}x^*-\min_\text{off}x^*\geq0, \\
 \delta_\text{on}(x^*) \equiv \max_\text{on}x^* - \min_\text{on}x^*\geq0, \\
 \delta_\text{off}^\text{on}(x^*) \equiv \min_\text{on}x^* - \max_\text{off}x^*\leq1.
 \label{eq:def_actdeg}
\end{aligned}
\end{equation} 
The definitions enable a straightforward comparison with an all-or-none switch that would take on the optimum values $\delta_\text{off}=\delta_\text{on}=0$, $\delta_\text{off}^\text{on}=1$ and would achieve the maximum quality $Q_\text{id}=1$. Therefore, the switch quality measures the ``distance'' between the actual realisation and the ideal switch. A higher quality corresponds to a larger difference between the activations in the ``off'' and ``on'' state and is associated with an increased robustness, which are both biologically relevant. 
For all following investigations of the switch quality, the maximum value $\max_\text{on}x^*$ is taken from $[I]=3$ as shown in \cref{fig:act+bi_one}(a). 
It can be seen from the data in \cref{fig:act+bi_one}(c) that the switch quality of the single-species switch does not depend strongly on the chemical affinity and slightly increases with increasing $\dm$ until it reaches its plateau value of $Q_x\simeq0.63$ at $\dm/\kb T=12$.

\subsubsection*{Comparison of stochastic and deterministic description}

\begin{figure}\centering
\includegraphics[width=0.5\textwidth]{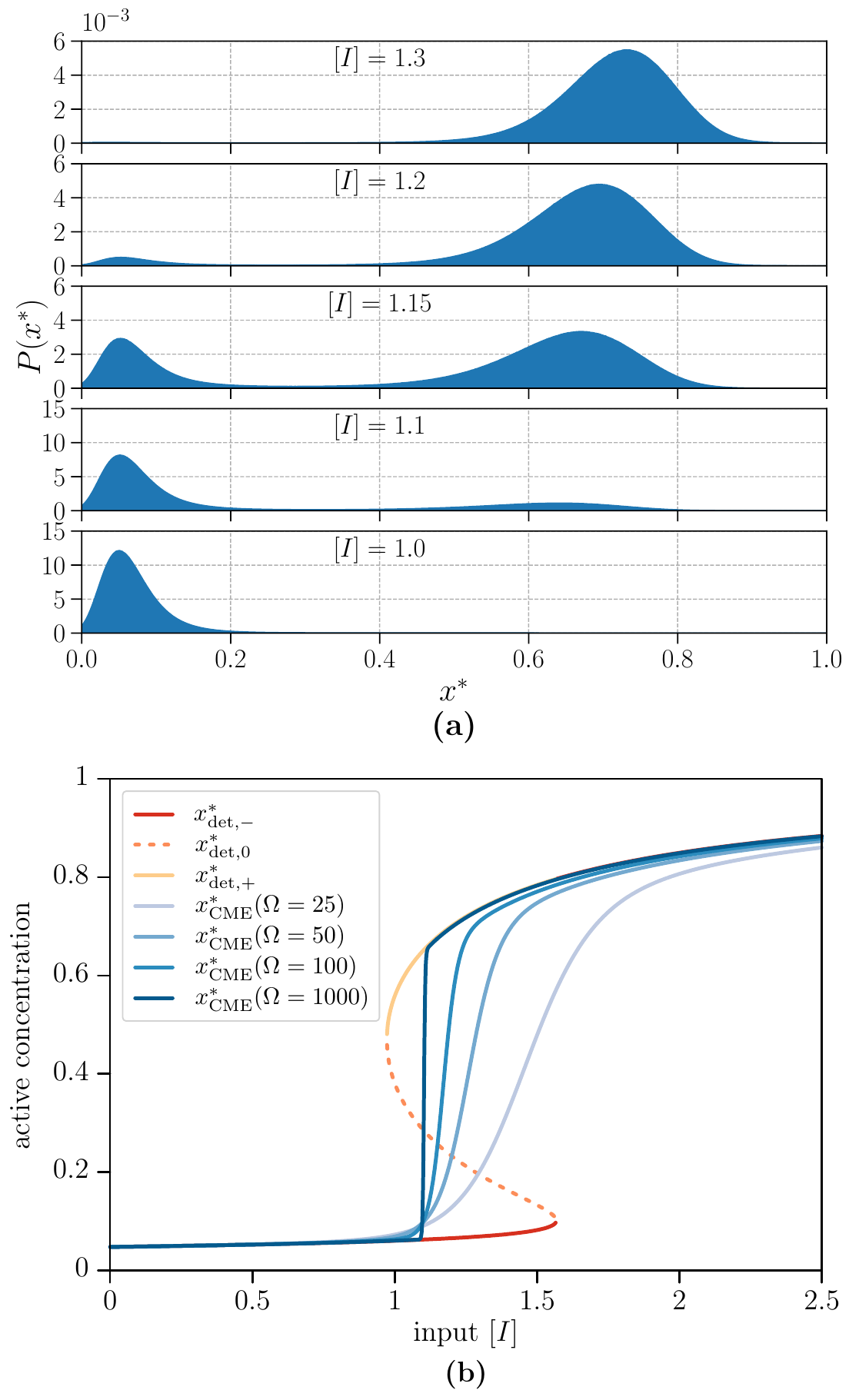}
 \caption{Probability distributions and active concentrations of the single-species switch. (a) Probability distribution $P(x^*)$ for different input concentrations $[I]$, chemical affinity $\dm/\kb T=10$, and system size $\Omega=100$. (b) Active concentrations $x^*_{\text{det},i}$ from the solutions of \cref{eq:det-one} in comparison with the average concentrations $x^*_\text{CME}$ of \cref{eq:concen_def} from the CME for different system sizes $\Omega$ as functions of the input $[I]$ for $\dm/\kb T=10$. Deterministic and stochastic concentrations are in good agreement for $\Omega\gtrsim1000$ apart from the metastable region.}
 \label{fig:prob+concen_one}
\end{figure}

In this section, we study the occupation of the two stable steady states in the bistability region and compare deterministic and stochastic descriptions. The probability distribution $P(x^*)$ is shown in \cref{fig:prob+concen_one}(a) for different input concentrations. 
In an experimental realisation of such switches, the protein concentration will stay in one of the bistable solutions for a very long time. In \cref{fig:prob+concen_one}(b), we compare the active concentrations $x^*_{\text{det},i}$ for the different solutions of \cref{eq:det-one} with an average concentration $x^*_\text{CME}$ computed from the CME via \cref{eq:concen_def}, which is plotted for different system sizes $\Omega=25,50,100,1000$. Increasing the system size has two major advantages. First, the concentration in the active state increases, as expected. Second, the transition region from the lower stable solution to the upper one, in which apparent bistability is observed, decreases with increasing system size. For a system size $\Omega=1000$, the average of the stochastic concentration is in good agreement with the deterministic concentrations apart from the metastable region and the apparent bistability region is small. Ge and Qian \cite{ge09} have shown that the chemical master equation converges to the deterministic solution with the lowest non-equilibrium potential. Note that the descriptions by the chemical master equation and the deterministic equation differ in the order in which the limits of time ($t\to\infty$) and volume ($\Omega\to\infty$) are taken. The CME considers the limits $\lim_{\Omega\to\infty}\lim_{t\to\infty}$, i.e., the time limit first and the volume limit second, whereas the deterministic equation takes the limits in opposite order, $\lim_{t\to\infty}\lim_{\Omega\to\infty}$.


\subsection*{Advantages of interlinked cascades over single-species switches}

\subsubsection*{Multiple non-equilibrium phases}

\begin{figure*}\centering
 \includegraphics[width=\textwidth]{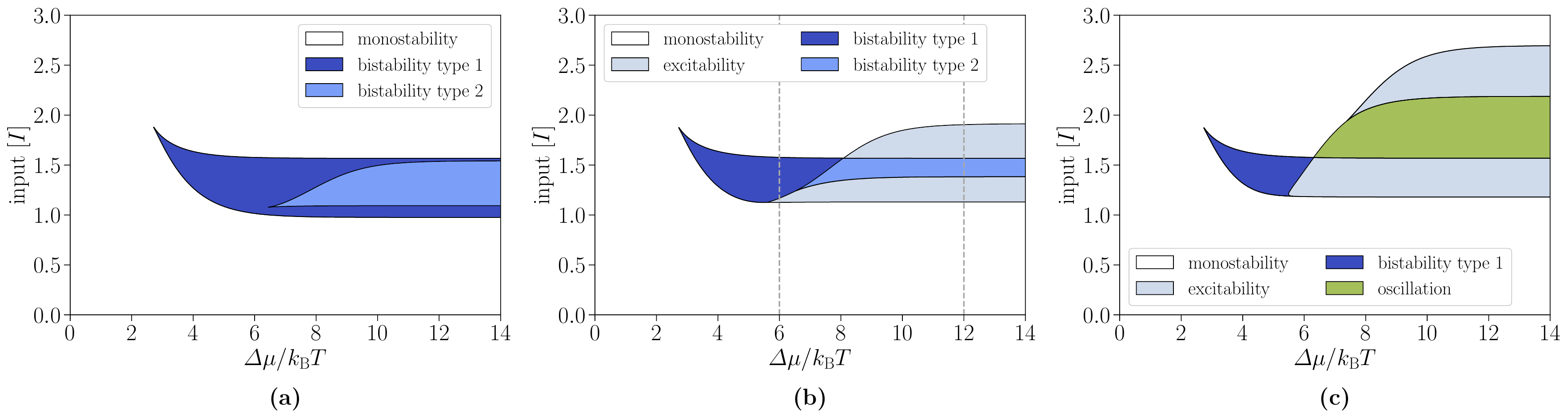}
 \caption{Phase diagrams of the interlinked cascade for different negative feedback strengths (a) $[G_1^-]=0.01$, (b) $[G_1^-]=1.0$, and (c) $[G_1^-]=3.0$. Increasing the negative feedback strength yields a transition from bistability to excitability and oscillation. The grey dashed lines indicate the values for the activation functions in \cref{fig:active_two}(a),(b). The other parameter values are specified in the Methods section. Note that the different non-equilibrium regimes correspond to both species $X$ and $Y$.}
 \label{fig:pd_two}
\end{figure*}

\begin{figure*}\centering
\includegraphics[width=\textwidth]{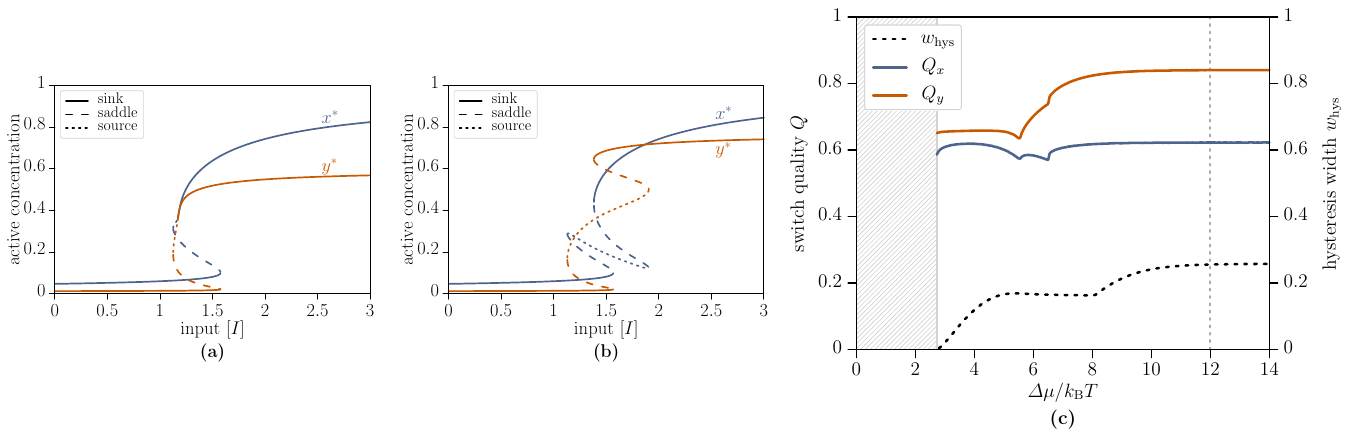}
 \caption{Activation functions and performance of the interlinked cascade. (a),(b) Active concentrations $x^*$ and $y^*$ as functions of the input $[I]$ for different chemical affinities (a) $\dm/\kb T=6$ and (b) $\dm/\kb T=12$, for the parameter values marked in \cref{fig:pd_two}(b). A higher chemical affinity $\dm$ leads to the emergence of excitability and a better switching behaviour of $y^*$. (c) Relative hysteresis width $w_\text{hys}$ from \cref{eq:w_hys} and switch qualities $Q_x$ and $Q_y$ from \cref{eq:def_quality} as functions of the chemical affinity $\dm$ for the interlinked cascade with $[G_1^-]=1.0$. If the chemical affinity is high enough, the second component ($Q_y$) has a much better performance than the first component ($Q_x$). The grey dashed lines indicate the values for the activation functions in (a),(b).}
 \label{fig:active_two}
\end{figure*}

The phase space of the interlinked cascade is much richer than the phase space of the single-species switch. While only two phases (monostability and bistability) exist for the single-species switch, the interlinked cascade is additionally able to exhibit oscillation, excitability and tristability. We provide a brief overview of the most relevant nonequilibrium phases in this main text, detailed information can be found in Supplementary Note 1.  Oscillation is characterised by the existence of a limit cycle, on which the trajectory circles around a source, which is the sole steady state. If a sink and a saddle arise additionally to the source, the phase is called excitability since the system can be excited to a quasi steady state. Additionally, a second type of bistability arises, which differs from the former bistability regime by an additional source and one further saddle.

The phase diagrams of the interlinked cascade for different negative feedback strengths $[G_1^-]=0.01,1.0,3.0$ are presented in \cref{fig:pd_two}. As a first result, there occur phase transitions from bistability to excitability and oscillation with increasing negative feedback strength from $[G_1^-]=0.01$ to $[G_1^-]=3.0$. Hence, by varying the experimentally accessible concentration of $G_1^-$ molecules, the network motif can be tuned to achieve both bistability and oscillations.
Second, the diagrams reveal that the chemical affinity switches between different non-equilibrium phases.

Moreover, our model is able to achieve regions of tristability, where three stable steady states coexist. A detailed analysis in Supplementary Note 2 reveals that tristability is not an additional regime of the interlinked cascade as the probability of the third stable steady state is very low, making an experimental observation impossible. Consequently, the behaviour of the tristable system is quite similar to the bistable one.

\subsubsection*{All-or-none activation}


\begin{table}\centering
  \caption{Comparison of the switch qualities $Q$ and hysteresis widths $w_\text{hys}$ for different network motifs at a chemical affinity of $\dm/\kb T=12$.}
  \begin{tabular}{llll}
  \toprule
  switch type &  & $Q$ & $w_\text{hys}$ \\
  \midrule
  all-or-none switch & & 1 & -- \\ 
  single-species switch & $x^*$ & 0.63 & 0.235 \\ 
  \multirow{2}{*}{interlinked cascade} & $x^*$ & 0.62 & \multirow{2}{*}{0.255} \\ 
  & $y^*$ & 0.84 &  \\
  \bottomrule
 \end{tabular}
 \label{tab:actdeg_comp}
\end{table}

The activating effect of the chemical affinity becomes apparent by studying the bifurcation diagrams of \cref{fig:active_two}(a),(b). In these figures, the active concentrations $x^*$ and $y^*$ are plotted as functions of the input $[I]$ for chemical affinities (a) $\dm/\kb T=6$ and (b) $\dm/\kb T=12$. The activation of the second component changes drastically with increased chemical affinity and leads to a better switching behaviour.

We quantify the above statement by studying the switch quality for the interlinked cascade. The qualities $Q_x$ and $Q_y$ of both components are shown in \cref{fig:active_two}(c) as function of the chemical affinity and allow for a direct comparison with the single-species switch. In contrast to the latter, the switch quality depends considerably on the chemical affinity for $\dm/\kb T\lesssim10$. For larger affinities, $Q_x$ and $Q_y$ approach their plateau values, which are presented in \cref{tab:actdeg_comp} together with the corresponding values for the single-species and the all-or-none switch.
It is apparent from these data that the second component of the interlinked cascade leads to a significantly better performance with a switch quality of $Q_y\simeq0.84$ at $\dm/\kb T=12$. The associated hysteresis width and the quality of the first component remain almost unaffected with $w_\text{hys}\simeq0.255$ and $Q_x\simeq0.62$. Besides, the data reveal the importance of the chemical affinity since the switch quality is substantially improved for values $\dm/\kb T\gtrsim7$ . Overall, the activation performance of the interlinked cascade is relatively close to the ideal all-or-none switch for high enough chemical affinities. 
The observed kinks in the switch qualities at $\dm/\kb T\simeq5.5,6.5$ as well as the kink in the relative hysteresis width at $\dm/\kb T\simeq8.1$ can be attributed to the onsets of the lower excitability region, bistability type 2, and the upper excitability region, respectively (see \cref{fig:pd_two}(b)).

\subsubsection*{Robustness against noisy input}

\begin{figure}\centering
 \includegraphics[width=0.5\textwidth]{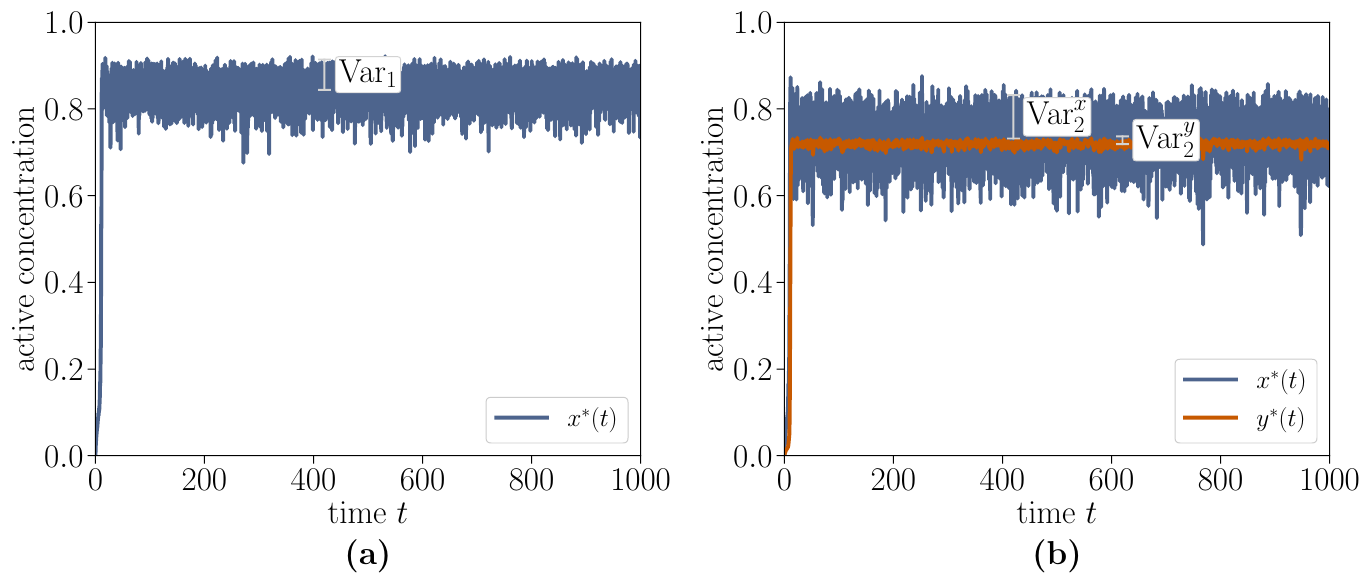}
 \caption{Time evolution of the active concentrations for noisy input of the single-species switch (a) and the interlinked cascade (b) for $[I_0]=2.0$ and $\dm/\kb T=12$. The second component of the interlinked cascade is significantly less noisy compared to the single-species switch as indicated by the variances. The remaining parameter values are a noise intensity of $D=0.005$, final integration time $t_\ff=10^3$, and initial concentrations $(x_0^*,y_0^*)=(0,0)$.}
 \label{fig:timeevol_noiseinp}
\end{figure}

\begin{figure}\centering
\includegraphics[width=0.5\textwidth]{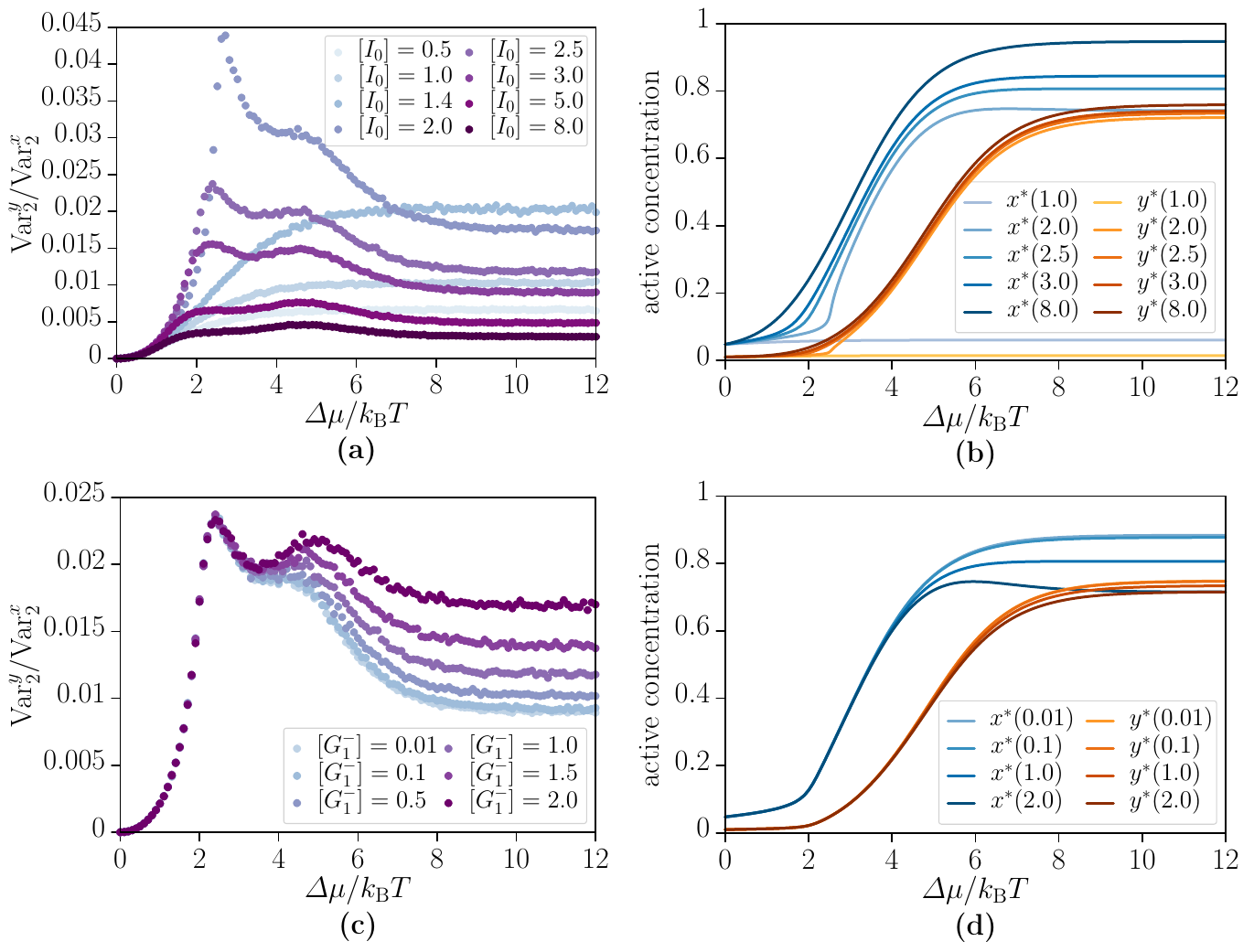}
 \caption{Variance ratio $\text{Var}_2^y/\text{Var}_2^x$ of the two components of the interlinked cascade as function of the chemical affinity $\dm$. (a),(b) Variance ratio for different input concentrations $[I_0]$ with fixed $[G_1^-]=1.0$ and associated concentrations. The ratio changes significantly for different input values. (c),(d) Variance ratio for different negative feedback strengths $[G_1^-]$ with fixed $[I_0]=2.5$ and associated concentrations. The variance ratio increases with increasing negative feedback. The noise intensity is $D=0.005$ and the initial concentrations are $(x_0^*,y_0^*)=(0,0)$ for all figures.} 
 \label{fig:variance}
\end{figure}

The results presented thus far provide insights into steady state properties of biological cascade switches. In this paragraph, we study the dynamical behaviour of the system. To this end, we consider a noisy input function to obtain a realistic biological signal \cite{tian09,zhan07}. The noisy input is described by
\begin{equation}
 [I(t)]=[I_0]\big(1+\xi(t)\big), 
 \label{eq:inp+noise}
\end{equation}
where $\xi(t)$ is normally distributed white noise with zero mean $\braket{\xi(t)}=0$ and correlations $\braket{\xi(t)\xi(t')}=2D\delta(t-t')$. Here, the parameter $D$ denotes the noise intensity and $\braket{\cdot}$ denotes an ensemble average. 
This stochastic input is used in the ordinary differential \cref{eq:det-one} for the single-species switch and \cref{eq:det-two} for the interlinked cascade.
In order to keep the analysis as simple as possible, we consider input fluctuations solely in the monostable regime. 
The response of the single-species switch and interlinked cascade to the noisy input is shown in \cref{fig:timeevol_noiseinp} for $[I_0]=2.0$, $D=0.005$, and $\dm/\kb T=12$. The fundamental difference between the two switches is that the second component of the interlinked cascade is much less noisy than the single-species switch. This enhanced robustness against fluctuations is a striking advantage of the interlinked cascade as it increases the distinction between active and inactive state by reducing the fluctuations. Hence, the interlinked cascade acts as a low-pass filter \cite{hoos05,zhan07}.

Having discussed the enhanced robustness qualitatively, we will now quantify our statements by computing the variances of the single-species switch ($\text{Var}_1$) and of the second component of the interlinked cascade ($\text{Var}_2^y$). The values for $[I_0]=2.0$, $D=0.005$, $t_\ff=10^6$, $\dt=0.01$, and $\dm/\kb T=12$ are
\begin{equation}
\begin{aligned}
 \frac{\text{Var}_1}{[I_0]\sqrt{2D/\dt}}\simeq4.0\cdot10^{-4}, && \frac{\text{Var}_2^y}{[I_0]\sqrt{2D/\dt}}\simeq1.85\cdot10^{-5},
\end{aligned}
\end{equation}
which leads to a variance ratio of $\text{Var}_2^y/\text{Var}_1\simeq0.046$. This small number reveals the enhanced robustness against noisy input of the second component which makes it easier for such biological switches to react to input signals.

We now turn to the behaviour of the variance ratio of the interlinked cascade $\text{Var}_2^y/\text{Var}_2^x$ as a function of the chemical affinity $\dm$. \Cref{fig:variance} provides the results obtained from simulations for varied input concentrations $[I_0]$ (top row) and negative feedback strengths $[G_1^-]$ (bottom row). Major differences between the input concentrations $[I_0]\leq1.4$ and $[I_0]\geq2.0$ can be seen from the data in \cref{fig:variance}(a) and (b). There is a clear trend of increasing variance ratio up to $[I_0]\simeq1.4$ along with almost no activation within this regime. For $[I_0]\simeq2.0$, two peaks emerge and the associated activation functions are remarkably stronger. Further increasing the input leads to a decrease of the peak heights and the variance ratio. The peak positions appear to correlate with the maximum increase of $x^*$ and $y^*$, which occur at different chemical affinity $\dm$.
The negative feedback strength $[G_1^-]$ has a strong impact on the variance ratio for $\dm/\kb T\gtrsim3$ (\cref{fig:variance}(c)). This is the most striking aspect of the data as it discloses a trade-off between robustness and oscillations. Increasing the negative feedback strength enables the interlinked cascade to perform oscillations, however, at the expense of a large variance ratio. Finally, increasing the chemical affinity above $\dm/\kb T\gtrsim6$ decreases the fluctuations, as recently observed in biochemical oscillators by \cite{junc18}.


\section*{Discussion and summary}\label{sec:disc}


In this study, we develop a thermodynamically consistent model for biochemical cascade switches that solely considers fundamental biochemical reactions. The use of trimolecular reactions may at first glance seem artificial. Nonetheless, our choice can be justified by the following reasons. First, we choose effective reactions as we are interested in a simple thermodynamically consistent network achieving both bistability and oscillations. It is well known that for an interlinked model as simple as ours, where the species have two states each, bistability or oscillations can only be achieved using trimolecular reactions if we assume mass action kinetics \cite{schl72, nova08}. In that sense, our interlinked cascade can be seen as a combination of Schlögl's model \cite{schl72} and the Brusselator \cite{nico77}. Second, it turns out that trimolecular reactions can be the result of several bimolecular reactions and additional species states. These states can then be coarse-grained into an active and inactive state using timescale separation, as shown for the Rho GTPase \cite{gory08}. In addition, minimal models with only bimolecular reactions that achieve bistability or oscillations have been identified \cite{wilh09}. However, they require additional species compared to their usual trimolecular counterparts, which is equivalent to having additional protein states.

Generally, switches must operate far from equilibrium ($\dm\neq0$) in order to observe interesting regimes such as bistability, excitability, and oscillation. Our model for a single-species switch is able to exhibit bistability if the chemical affinity is high enough. Bistability makes the switch robust against input fluctuations and is accompanied by a hysteresis behaviour. As a measure of the performance of switches, we define the switch quality using the activation degrees in the ''off`` and ''on`` states as well as in the transition between them. An ideal all-or-none switch corresponds to the maximum switch quality of 1 and exhibits hysteresis. We show that the quality of the single-species switch is considerably lower than the optimum and slightly increases with higher chemical affinity.

Adding a second species and connecting them with positive and negative feedback relations  leads to a more sophisticated model. The phase space of the interlinked cascade is richer and involves additional regimes, such as excitability, oscillation, and tristability. The interlinked cascade has several advantages over the single-species switch. First, the switch quality of the interlinked cascade is much higher than the one of the single-species switch and is closer to an ideal all-or-none switch. If we consider an input signal with noise, the interlinked cascade is shown to be much less noisy. Hence, it has an increased robustness against fluctuating signals and the distinction between the ''off`` and ''on`` state becomes clearer. Berg {\sl et al.} have also shown that fluctuations decrease the sharpness of the activation curve, providing further evidence that the interlinked cascade is a better motif for biochemical switches \cite{berg00}. The evidence presented in this study suggests that there exists a trade-off between switch quality, hysteresis width, and chemical affinity, which has to be balanced in biochemical switches.

The performance of the cascade could be further increased by adding more species to it. However, additional species may lead to a decrease of robustness as there is an optimal cascade length \cite{detw00,that02}. For small GTPases, interlinked cascades have only been identified between two species so far and regarding three species, merely GEF cascades (without negative feedback) have been discovered \cite{mizu12}.
There is abundant room for further progress in examining switches with different feedback structures, see for example \cite{kim06,kim08,tsai08}. Since these studies do not assume mass action kinetics, they lack thermodynamic consistency. We have explored variations of the interlinked cascade \cite{ehrm18}. By replacing the negative feedback loop with a positive one, we obtain another thermodynamically consistent switch. We can still achieve bistability and tristability, but as a drawback of this motif, the regimes of excitability and oscillation are no longer accessible. In case of a double negative feedback motif, it is very difficult to find a biologically relevant parameter range and most of the settings lead to an irreversible activation curve. These feedback structures could be relevant for systems with complex activation functions like gene expression \cite{tian09}. Nonetheless, our simple model presented here is sufficient to explain the versatility of interlinked small GTPases, which solely comprise positive-negative feedback structures.

Small GTPases have cofactors that are essential to achieve bistability or oscillations. Our model indicates that increasing the negative feedback, i.e., the GAP concentration, converts biochemical switches into oscillators. It would be very interesting to verify this prediction experimentally. Furthermore, a future study investigating how small GTPases respond to a heat shock, where the concentration ratio of GTP over GDP decreases, would be interesting \cite{trau94}. Besides, we expect the catalytic power of the cofactors to vary during stress to maintain the function of the cascade, as observed in the Hsp70 system \cite{nguy17}. An issue that was not addressed in this study is the membrane-cytoplasmic reaction and a third cofactor, namely the guanosine nucleotide dissociation inhibitor (GDI). It binds to the inactive GDP-bound state and transports it across the membrane \cite{mizu12,witt14a,witt14b}. From a dynamical perspective, the GDIs can further inactivate the system in the absence of feedback and can control the total concentration of active species. Assuming fast kinetics \cite{delc08}, we do not expect major qualitative differences when including GDIs to our model.

We emphasise that our model is, to the best of our knowledge, a simple thermodynamically consistent model that demonstrates the biological functions of interlinked cascades, such as bistability and oscillations. Although the focus of this study is on GTPase cascades, our results are general and can be applied to a broader class of systems. In a future step, we could incorporate our cascade into a larger signalling network to investigate how the switch quality affects biological processes.


\section*{Methods} \label{sec:methods}

The explicit values for the parameters of the single-species switch and interlinked cascade are chosen in such a way to satisfy biological requirements as well as to get reasonable bifurcation and phase diagrams. While most of the parameters are given in natural units ($\unit[1.0]{s^{-1}}$), the rate $\kappa_{22}^+$ must be larger to observe interesting non-equilibrium regions. However, it must be smaller than $\unit[4.0]{s^{-1}}$ to achieve reversible activation curves. We choose $\kappa_{22}^+=\unit[3.9]{s^{-1}}$ to enlarge the nonmonostable region and we take $\kappa_{11}^+=\kappa_{22}^+$ to keep the system as simple as possible. 
The rates $\omega_1^-$ and $\omega_2^-$ must be smaller than their counterparts $\omega_1^+$ and $\omega_2^+$ in order to achieve a dominant inactivation in the hydrolysis reactions. The transition rates are in the same order of magnitude as in \cite{pfeu09}. We set the parameters to $\omega_1^+=\unit[1.0]{s^{-1}}$, $\omega_1^-=\unit[0.05]{s^{-1}}$, $\omega_2^+=\unit[1.0]{s^{-1}}$, $\omega_2^-=\unit[0.01]{s^{-1}}$, $\kappa_{11}^+=\unit[3.9]{s^{-1}}$, $\kappa_{22}^+=\unit[3.9]{s^{-1}}$, $\kappa_{12}^+=\unit[1.0]{s^{-1}}$, $\omega_{21}^+=\unit[1.0]{s^{-1}}$, $[G_1^-]=1.0$, $[G_2^+]=1.0$, $D=0.005$, $\dt=0.01$. The remaining four reaction rates $\kappa_{11}^-$, $\kappa_{22}^-$, $\kappa_{12}^-$, $\omega_{21}^-$ are fixed by the local-detailed balance conditions of \cref{eq:ldb_two}. Note that the negative feedback strength $[G_1^-]$ is changed in \cref{fig:pd_two,fig:variance}.

The statics and dynamics of the system are analysed by deterministic mass action kinetics as well as chemical master equations and chemical Langevin equations. 
A fourth order Runge-Kutta method \cite{stro00} is used to solve the ordinary differential equations numerically with an integration time step $\dt=0.01$ and a final integration time $t_\ff$.
We used Mathematica (Wolfram Research, Inc., Champaign, Illinois), Python (Python Software Foundation), and C++ (Standard C++ Foundation) for the simulations.




\subsection*{Data accessibility}
All data essential to evaluate and establish the conclusions of this study are presented in the paper itself. Computer codes can be obtained from the author upon request.

\subsection*{Author contributions}
A.E. performed research. A.E., B.N., and U.S. designed research, discussed the results and wrote the paper.

\subsection*{Competing interests}
The authors declare no conflict of interest.

%


\clearpage
\newpage

\onecolumngrid
\section*{\large \textbf{Supplementary Information}} 
\begin{center}
\textbf{Interlinked GTPase cascades provide a motif for both robust switches and oscillators}\\ \vspace{0.5cm}
  Andreas Ehrmann, Basile Nguyen, and Udo Seifert\\
  \emph{II. Institut für Theoretische Physik, Universität Stuttgart, 70550
    Stuttgart, Germany}\\
  (Dated: \today)
\end{center}

\setcounter{equation}{0}
\setcounter{figure}{0}
\setcounter{table}{0}
\setcounter{page}{1}
\renewcommand{\theequation}{S\arabic{equation}}
\renewcommand{\thefigure}{S\arabic{figure}}
\renewcommand{\bibnumfmt}[1]{[S#1]}
\renewcommand{\citenumfont}[1]{S#1}

\section*{Supplementary Note 1. Linear stability and phase plane analysis}\label{sec:ppa-lsa}


\subsection*{Linear stability analysis}\label{subsec:lsa}

The stability of the fixed points of the system is determined by a linear stability analysis in which we consider the following set of non-linear differential equations
\begin{equation}
 \frac{\dd\bm{x}}{\dt} = \bm{f}(\bm{x}),
\end{equation}
with the concentration vector $\bm{x}=(x_1,\dots,x_M)^T$ of $M$ different protein species and the equation vector $\bm{f}(\bm{x})=(f_1(\bm{x}),\dots,f_M(\bm{x}))^T$.
In linear stability analysis, such non-linear systems are approximated around each fixed point $\bm{x}_0$ by a linear system via
\begin{equation}
 \bm{f}(\bm{x}) \approx \bm{f}(\bm{x}_0) + \mathbf{J}(\bm{x}_0)(\bm{x}-\bm{x}_0).
\end{equation}
To this end, one determines the fixed points $\bm{x}_0$ by solving $\bm{f}(\bm{x}_0)=0$, computes the Jacobian $\mathbf{J}(\bm{x}) \equiv \partial_{\bm{x}}\bm{f}(\bm{x})$ and its eigenvalues $\lambda_i$ at each fixed point $\bm{x}_0$. The stability of each fixed point is based on the real parts of its eigenvalues. 

In one dimension, as for the single-species switch in Eq. 9 of the main text, the analysis is relatively simple and the stability of the fixed points is represented by the potential $U(x)=-\partial_{x}f_1(x)$. For two-dimensional systems as in Eq. 12 of the main text for the interlinked cascade, stable fixed points arise if both eigenvalues have negative real parts $\re(\lambda_1) < \re(\lambda_2) < 0$ and are called sinks. In all other cases, the fixed point is unstable. For one positive and one negative value $\re(\lambda_1) < 0 < \re(\lambda_2)$, it is called saddle. For a source, both are positive $\re(\lambda_1) > \re(\lambda_2) > 0$.
The entries of the Jacobian for the interlinked cascade are given by
\begin{equation}
\begin{aligned}
 \partial_xf_1(x,y) &= -\omega_1^- - \omega_1^+ + 2[I]\kappa_{11}^+x - 3[I](\kappa_{11}^++\kappa_{11}^-)x^2 - [G_1^-](\omega_{21}^-+\omega_{21}^+)y^2,\\
 \partial_yf_1(x,y) &= 2[G_1^-]\omega_{21}^-(1-x)y - 2[G_1^-]\omega_{21}^+xy, \\
 \partial_xf_2(x,y) &= 2[G_2^+]\kappa_{12}^+x(1-y) - 2[G_2^+]\kappa_{12}^-xy, \\
 \partial_yf_2(x,y) &= -\omega_2^- -\omega_2^+ - [G_2^+](\kappa_{12}^-+\kappa_{12}^+)x^2 + 2[G_2^+]\kappa_{22}^+y - 3[G_2^+](\kappa_{22}^+ + \kappa_{22}^-)y^2. 
\end{aligned}
\end{equation}


\subsection*{Phase plane analysis of the interlinked cascade}

\begin{figure*}\centering
 \includegraphics[width=0.6\textwidth]{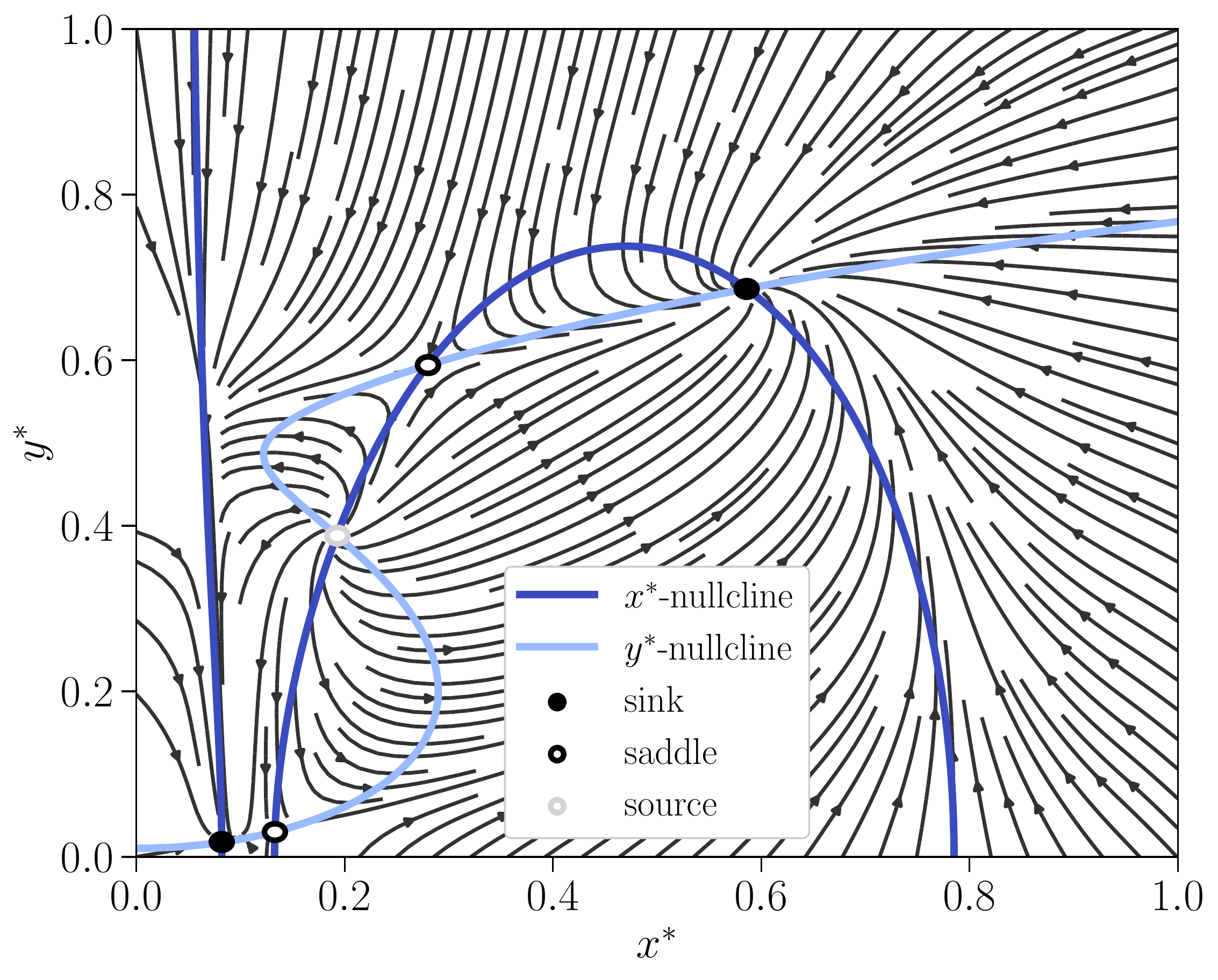}
 \caption{Phase plane in the bistability type 2 regime with parameter values $[I]=1.5$ and $\dm/\kb T=12$. The $x^*$- and $y^*$-nullclines are drawn in blue, the fixed points of the ODEs (Eq. 12 of the main text) are represented by the circles, and the contour lines of the ODEs are depicted by the anthracite arrows.}
 \label{fig:phaseplane}
\end{figure*}

As the coupled non-linear ODEs (Eq. 12 of the main text) are not analytically solvable, we perform a phase plane analysis to examine the fixed points of the system and determine all possible classes of dynamical behaviour through an exhaustive search over parameters. 
In \cref{fig:phaseplane}, a typical phase plane is presented with parameter values $[I]=1.5$ and $\dm/\kb T=12$. The two nullclines $\dd x^*/\dt=0$ and $\dd y^*/\dt=0$  are represented by the blue lines and the circles represent its non-equilibrium steady states. The two stable fixed points are marked as black filled circles, while the three unstable fixed points are marked as black open circles for the saddles and a grey open circle for the source. 
When the input $[I]$ is varied, the $x^*$-nullcline primarily changes, whereas the $y^*$-nullcline is nearly unaffected. 

The number of sinks $l$ (stable), saddles $m$ (unstable) and sources $n$ (unstable) characterise the different non-equilibrium phases of the system, which are denoted by a triplet $(lmn)$. 
Seven different phases can be identified by an exhaustive search over parameters: monostability (100), bistability type 1 (210), bistability type 2 (221), excitability (111), oscillation (001), tristability type 1 (320), and tristability type 2 (331).
These phases are connected by three types of bifurcations, namely saddle-node, Hopf, and saddle-node infinite period (SNIPER) bifurcations. A saddle-node bifurcation occurs when the system goes through a bistability region. A Hopf bifurcation occurs when the system performs a transition between a monostability (100) and oscillation (001) region and a SNIPER bifurcation occurs in case of a transition between an excitability (111) and oscillation (001) region. 

\begin{figure*}\centering
\includegraphics[width=0.9\textwidth]{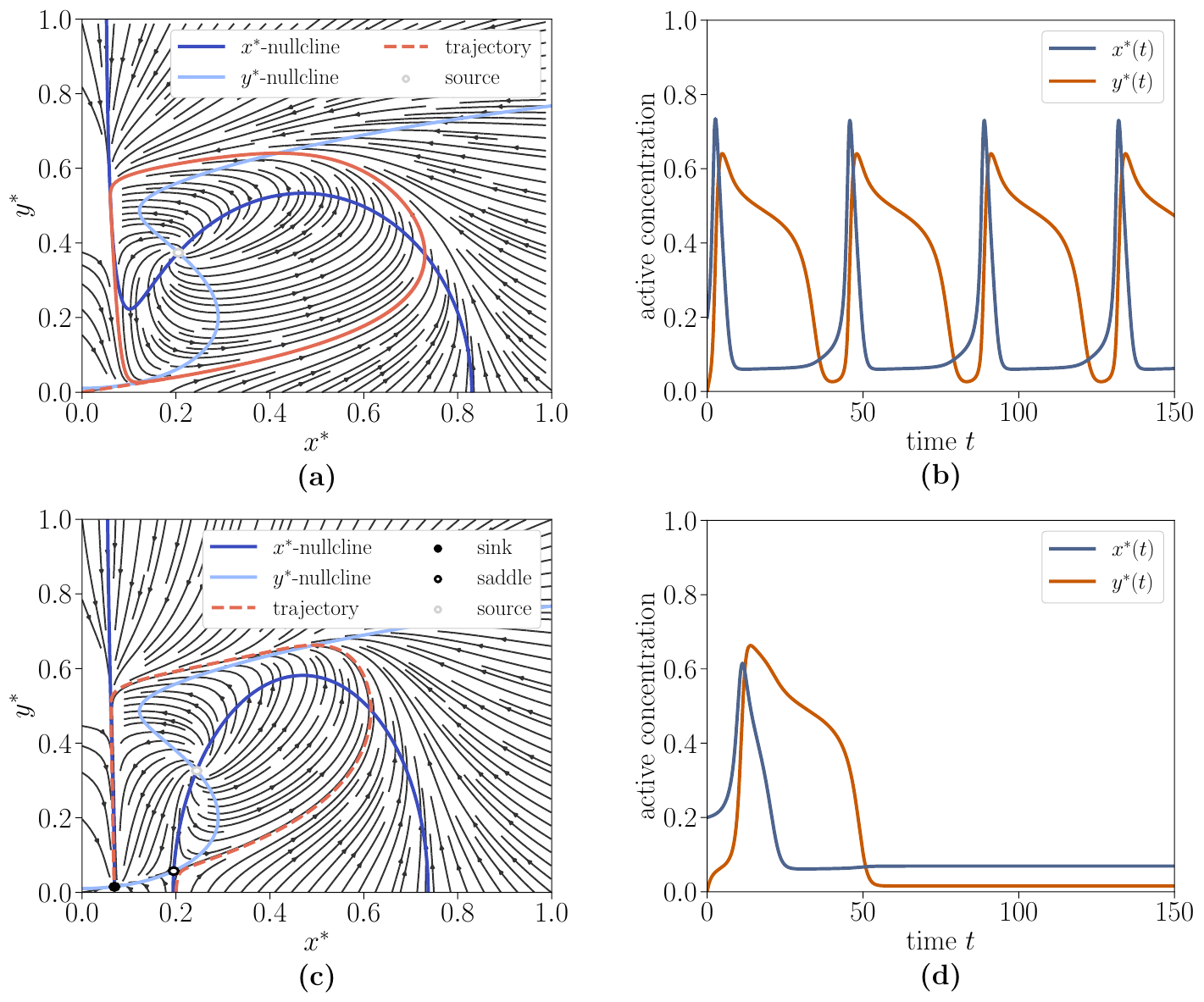}
 \caption{Phase planes and associated time evolutions of the active concentrations for constant input in the oscillation and excitability regime. (a),(b) Oscillation regime, in which the system performs oscillations along the limit cycle, with parameters $[I]=1.8,\dm/\kb T=12$, and $[G_1^-]=3.0$. (c),(d) Excitability regime, in which the system can be excited to a quasi-stationary manifold, with parameters $[I]=1.3,\dm/\kb T=12$, and $[G_1^-]=1.0$.}
 \label{fig:pp_two_exc+osc}
\end{figure*}

Oscillation is characterised by the existence of a limit cycle, on which the trajectory circles around the source, which is the sole steady state. This is shown in \cref{fig:pp_two_exc+osc}(a),(b). If a sink and a saddle arise additionally to the source, the phase is called excitability due to its activation properties. If the initial concentrations $(x_0^*,y_0^*)$ are beyond the sink's attraction region, the system can be excited to a quasi-stationary manifold. As presented in \cref{fig:pp_two_exc+osc}(c),(d) the trajectory follows the unstable manifold of the saddle and then returns along the stable manifold of the sink. The trajectory in the excitable regime looks very similar to one oscillation phase of the trajectory in the oscillatory regime. The driving mechanism of this behaviour could be a heteroclinic orbit which becomes a limit cycle after the SNIPER bifurcation. 
Additionally, a second type of bistability arises, which differs from the former bistability regime by an additional source and one further saddle. The same holds for the two different tristability regimes and the phases with a source and an additional saddle are called type 2, while the others are called type 1. 


\section*{Supplementary Note 2. Examining tristability with the effective potential}\label{sec:pot+tri}

For two-dimensional systems, it is not possible to calculate an explicit potential function $U(x,y)$ because our interlinked model is not a gradient system, meaning that $\partial_yf_1(x,y)\neq \partial_xf_2(x,y)$ \cite{verd14,stro00b}. 
However, an effective potential can be calculated according to the steady state probability distribution
\begin{equation}
 U(\bm{x}) = -\ln P_\st(\bm{x}).
\end{equation}
We consider a stochastic version of the system in order to determine the probability distribution by studying the chemical Langevin equations of the interlinked cascade, which are given by \cite{gill00}
\begin{equation}
\begin{aligned}  
 \frac{\dd x^*}{\dt} &= \left(\omega_1^- + [I]\kappa_{11}^+{x^*}^2 + [G_1^-]\omega_{21}^- {y^*}^2\right)(1-x^*)-\left(\omega_1^+ + [I]\kappa_{11}^-{x^*}^2 + [G_1^-]\omega_{21}^+ {y^*}^2\right)x^*\\
 &\quad+ \Omega^{-1/2}\left(\sqrt{\omega_1^-(1-x^*)}\xi_1 + \sqrt{[I]\kappa_{11}^+{x^*}^2(1-x^*)}\xi_2 + \sqrt{[G_1^-]\omega_{21}^- {y^*}^2(1-x^*)}\xi_3\right) \\
 &\quad- \Omega^{-1/2}\left(\sqrt{\omega_1^+x^*}\xi_4 + \sqrt{[I]\kappa_{11}^-{x^*}^3}\xi_5 + \sqrt{[G_1^-]\omega_{21}^+{y^*}^2x^*}\xi_6 \right), \\ 
 \frac{\dd y^*}{\dt} &= \left(\omega_2^- + [G_2^+]\kappa_{22}^+{y^*}^2 + [G_2^+]\kappa_{12}^+{x^*}^2\right)(1-y^*) - \left(\omega_2^+ + [G_2^+]\kappa_{22}^-{y^*}^2 + [G_2^+]\kappa_{12}^-{x^*}^2\right)y^* \\ 
 &\quad+ \Omega^{-1/2}\left(\sqrt{\omega_2^-(1-y^*)}\xi_7 + \sqrt{[G_2^+]\kappa_{22}^+{y^*}^2(1-y^*)}\xi_8 + \sqrt{[G_2^+]\kappa_{12}^+ {x^*}^2(1-y^*)}\xi_9\right) \\ 
 &\quad- \Omega^{-1/2}\left(\sqrt{\omega_2^+y^*}\xi_{10} + \sqrt{[G_2^+]\kappa_{22}^-{y^*}^3}\xi_{11} + \sqrt{[G_2^+]\kappa_{12}^-{x^*}^2y^*}\xi_{12} \right), 
 \label{eq:cleq}
\end{aligned} 
\end{equation}
with independent Gaussian white noise
\begin{equation}
\begin{aligned}
 \braket{\xi_i(t)}=0, && \braket{\xi_i(t)\xi_j(t')}=2D\delta_{ij}\delta(t-t'), 
\end{aligned}
\end{equation}
where $i,j\in\{1,\dots,12\}$.
We omit the explicit time dependence of $x^*(t)$, $y^*(t)$, and $\xi_i(t)$ in the ODEs for simplicity reasons.
In the simulations, we set the noise intensity to $D=0.005$. Note that in the thermodynamic limit $\Omega\to\infty$, the deterministic equation 12 of the main text is recovered.

\begin{figure*}\centering
\includegraphics[width=0.9\textwidth]{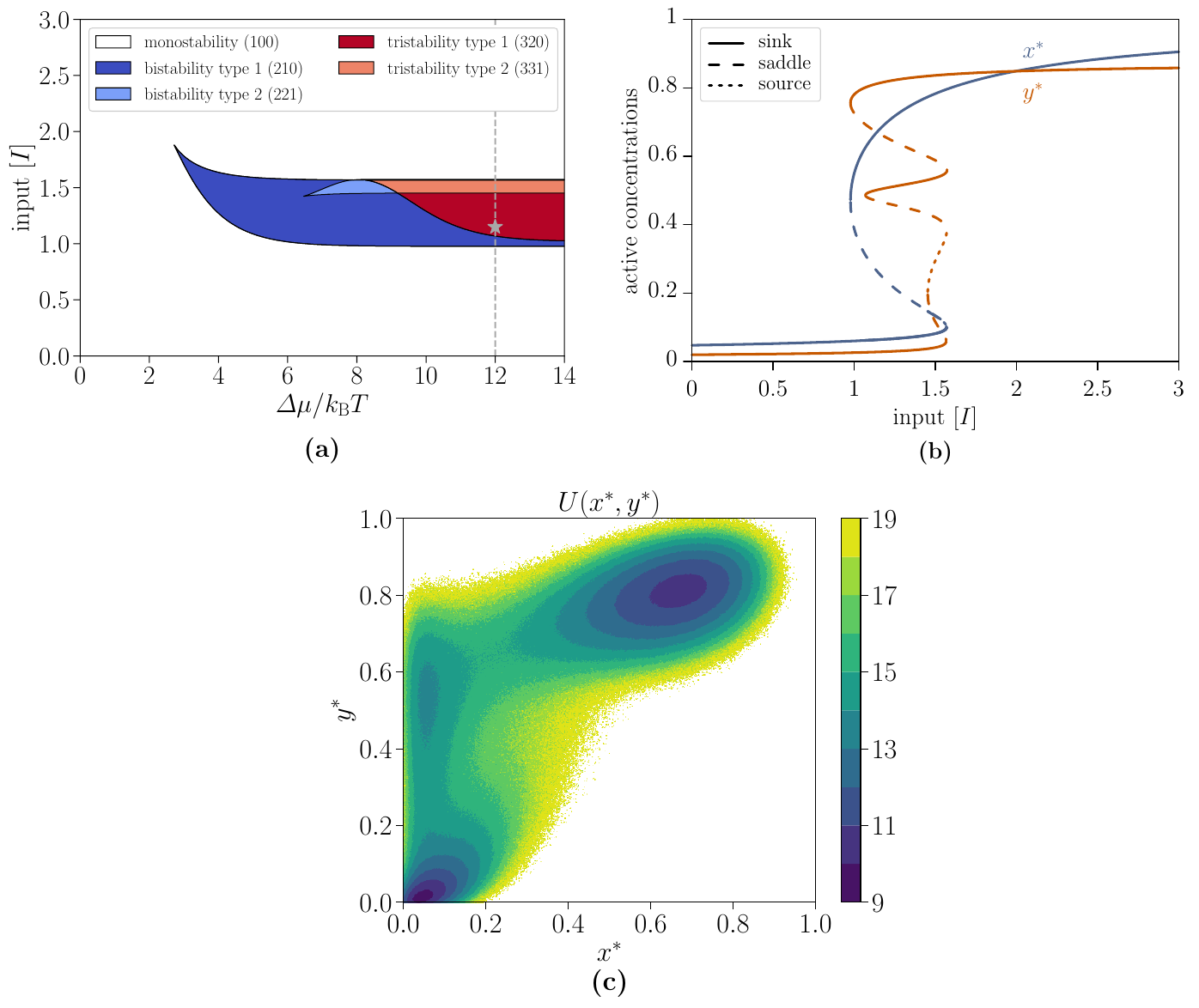}
 \caption{Tristability regime of the interlinked cascade. (a) Phase diagram of the interlinked cascade for negative feedback strength $[G_1^-]=0.01$ and reaction rate $\kappa_{12}^+=\unit[3.9]{s^{-1}}$ showing tristability. The grey line and star mark the parameters used in figures (b) and (c), respectively. (b) Activation function in the tristability regime for $\dm/\kb T=12$. (c) Effective potential $U(x^*,y^*) = -\ln P_\st(x^*,y^*)$ determined by the chemical Langevin \cref{eq:cleq} for the tristability region with $[I]=1.15$. The potential of the upper left NESS is four levels smaller than the potential of the other two.}
 \label{fig:tri}
\end{figure*}

By increasing the reaction rate from $\kappa_{12}^+=\unit[1.0]{s^{-1}}$ in Fig. 5(a) of the main text to $\kappa_{12}^+=\unit[3.9]{s^{-1}}$, the new regimes of tristability type 1 and type 2 are observable, which is shown in the phase diagram of \cref{fig:tri}(a). The associated activation function for $\dm/\kb T=12$ is shown in \cref{fig:tri}(b). Note that the switch quality $Q_y$ is even better as for the regime studied in the main text (if we neglect the improbable middle steady state). This supports our statements about the improved performance of the interlinked cascade. 
Theoretically, the system is able to exhibit tristability for this parameter set. In order to examine the occupation probabilities of these three stable steady states, the effective potential $U(x^*,y^*) = -\ln P_\st(x^*,y^*)$ determined by the chemical Langevin \cref{eq:cleq} is studied. As presented in the contour plot in \cref{fig:tri}(c), the potential of the third NESS in the top left corner is four levels higher than the potential of the other two. Correspondingly, the probability is four orders of magnitude smaller, making an experimental observation impossible. Hence, the behaviour of the tristable system is quite similar to the bistable one.
Note that the first component in \cref{fig:tri}(b) exhibits tristability in the same range as the second component although there is no evidence in the figure as the additional stable and unstable steady states have the same $x^*$-value at the lower stable steady state as shown in \cref{fig:tri}(c).

We emphasise that the statements in this paragraph hold for all parameter regimes studied in this paper. However, the existence of another parameter range, in which genuine tristability may occur is conceivable but is not an aim of this research.

\clearpage

\end{document}